\documentclass{article}

\usepackage{PRIME}

\usepackage[utf8]{inputenc} 
\usepackage[T1]{fontenc}    
\usepackage{hyperref}       
\usepackage{url}            
\usepackage{booktabs}       
\usepackage{amsfonts}       
\usepackage{nicefrac}       
\usepackage{microtype}      
\usepackage{lipsum}
\usepackage{amsmath}
\usepackage{enumitem}
\usepackage{amssymb}
\usepackage{mathtools}
\usepackage{fancyhdr}       
\usepackage{graphicx}       
\usepackage[sorting=nyt,style=apa]{biblatex}
\title{Leveraging cough sounds to optimize chest x-ray usage in low-resource settings
}

\author{
Alexander Philip \\
Christian Medical Center \& Hospital, Purnia, Bihar, 854301, India
\AND
Sanya Chawla\\
Qure.ai Technologies, Mumbai 400059, India \\
\AND
  Lola Jover, George P. Kafentzis, Joe Brew \\
  Hyfe Inc. Wilmington, Delaware, 19801, USA
  \AND
  Vishakh Saraf, Shibu Vijayan\\
  Qure.ai Technologies, Mumbai 400059, India \\
  \AND
  Peter M. Small\\
  University of Washington, Department of Global Health, Seattle, 98105, U.S.A
  \AND
  Carlos Chaccour \\
  ISGlobal, Barcelona Institute for Global Health, Barcelona, 08036, Spain \\
  University of Navarra, Pamplona, 31008, Spain \\
  Centro de Investigación Biomédica en Red de Enfermedades Infecciosas, Madrid, Spain
}

\begin{document}
\maketitle

\begin{abstract}
Chest X-ray is a commonly used tool during triage, diagnosis and management of respiratory diseases. In resource-constricted settings, optimizing this resource can lead to valuable cost savings for the health care system and the patients as well as to and improvement in consult time. We used prospectively-collected data from 137 patients referred for chest X-ray at the Christian Medical Center and Hospital (CMCH) in Purnia, Bihar, India. Each patient provided at least five coughs while awaiting radiography. Collected cough sounds were analyzed using acoustic AI methods. Cross-validation was done on temporal and spectral features on the cough sounds of each patient. Features were summarized using standard statistical approaches. Three models were developed, tested and compared in their capacity to predict an abnormal result in the chest X-ray.
All three methods yielded models that could discriminate to some extent between normal and abnormal with the logistic regression performing best with an area under the receiver operating characteristic curves ranging from 0.7 to 0.78. Despite limitations and its relatively small sample size, this study shows that AI-enabled algorithms can use cough sounds to predict which individuals presenting for chest radiographic examination will have a normal or abnormal results. These results call for expanding this research given the potential optimization of limited health care resources in low- and middle-income countries.

\end{abstract}


\section{Introduction}
\label{sec:introduction}
Chest radiography (CXR) [1] is a valuable tool for the identification and management of pulmonary pathologies. However, knowing which patients have a sufficiently high pretest probability to justify the inconvenience and expense of radiography is complicated and is generally based on expert clinical decision-making informed by a history and physical exam [2]. An inexpensive and widely available triage tool that could identify patients who are most likely to have radiographic abnormalities would improve the efficiency of CXR utilization and improve the specificity of the diagnostic algorithm. Cough has long been used as a marker of lung disease.  Traditionally, the assessment of cough has relied on subjective evaluations by healthcare professionals and patients. Recently, objective and automated cough detection has been made possible by advances in acoustic artificial intelligence (AI)/machine learning (ML) techniques [3]. These techniques can extract diagnostically relevant features from cough signals, such as duration, intensity, frequency, as well as characteristic sound and temporal patterns. AI models are being tested to discern patterns associated with specific respiratory illness,  including chronic obstructive pulmonary disease, tuberculosis, Covid-19, and pneumonia. [3-7]  We hypothesized that acoustic analysis of solicited cough sounds using an AI model could help to predict the likelihood that a patient will have a normal or abnormal CXR. Here we report the feasibility and preliminary performance of an AI-enabled cough classifier’s ability to identify which patients are most likely to have abnormal CXRs based on cough sound.


\section{Methods}

\subsection{Patient Enrollment and Data Capture}
Between January 10 and March 12, 2023, all patients referred to the outpatient facility of Christian Medical Center and Hospital (CMCH) in Purnia, Bihar, India for chest radiography were offered enrollment in the study. The CXR referral was based on the clinician's judgment. Each patient was informed of the study and gave verbal consent before being moved to a well-ventilated room dedicated for cough collection. In this room, the data collector gave a detailed explanation of the study purpose, protocol and obtained written informed consent. The room was equipped with a desktop computer attached to a microphone. They were asked to cough at least five times without a mask opposite an open window facing away from the microphone. Coughs were initially sampled at $44100$ Hz but downsampled to $16000$ Hz using a polyphase implementation. The duration of each cough sound was $500$ ms.

\subsection{Digital Chest X-ray Interpretation using a Validated AI System}
Digital chest radiography was performed using the Agfa CR-10X Computer Radiography X-Ray machine. Images were interpreted as normal or abnormal with Qure.ai's AI-based CXR interpretation tool (QXR v $3.2.9$) which has been amply validated across several settings. This classification was used as the ground truth [8-10].

\subsection{Cough Analysis}
\begin{figure}
    \centering
    \includegraphics[scale=.6]{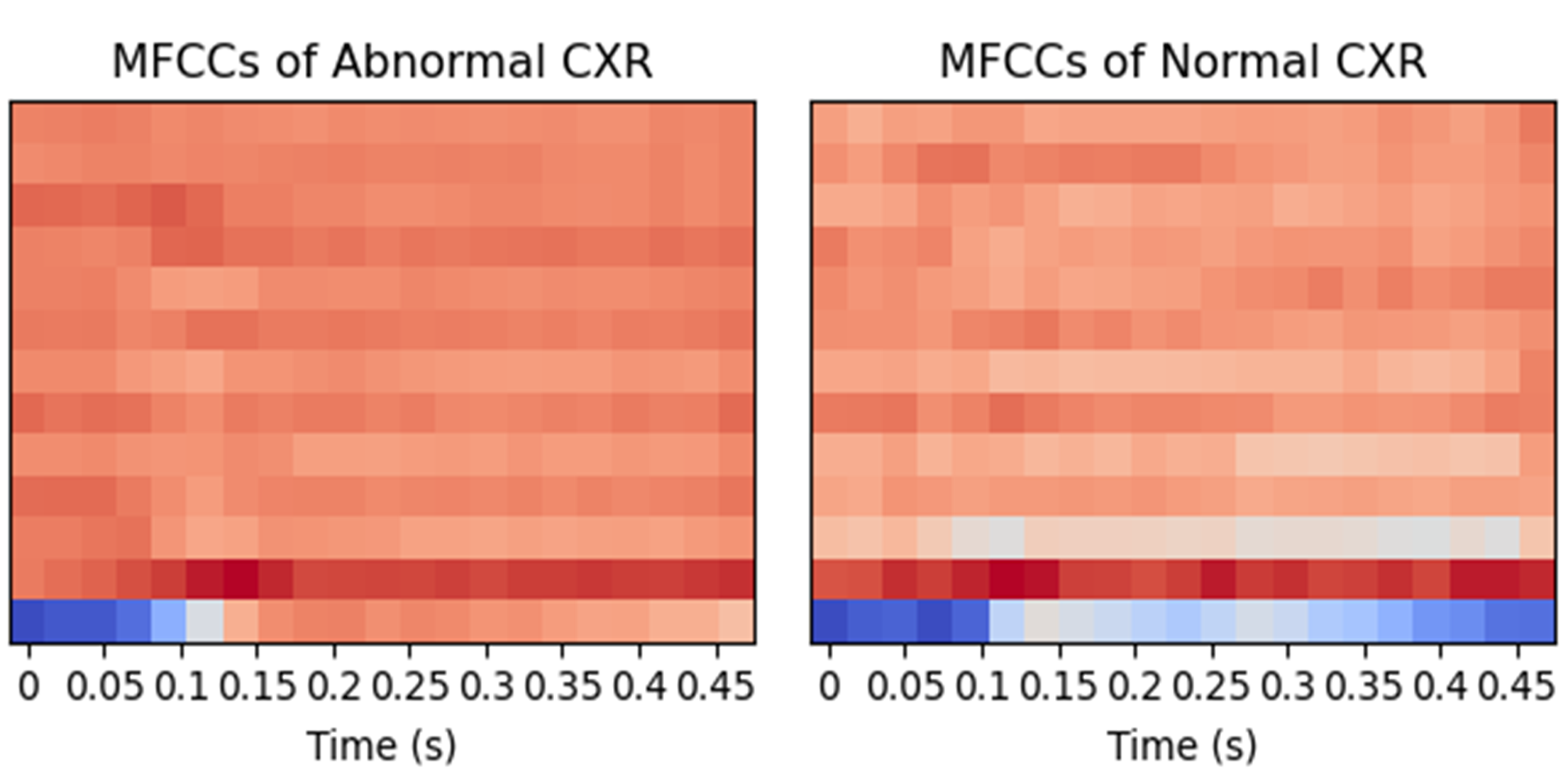}
    \caption{Mel Frequency Cepstral Coefficients of coughs from patients with (left) abnormal CXR and (right) normal CXR.}
    \label{fig:mfcc}
\end{figure}

\begin{figure}
    \centering
    \includegraphics[scale=.6]{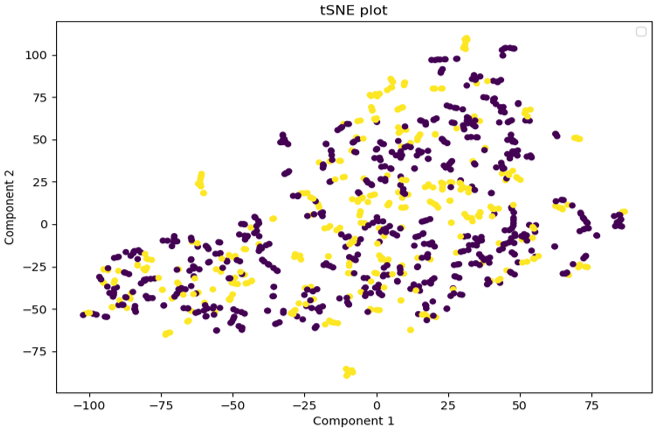}
    \caption{Two-dimensional tSNE plot of feature vectors. Yellow dots correspond to individuals with abnormal Xray while purple dots correspond to the ones with normal Xray.}
    \label{fig:tsne}
\end{figure}

Collected cough sounds were analyzed using acoustic AI methods. In brief, cross-validation was done on temporal and spectral features [11] on the cough sounds of each patient in a frame-by-frame manner (that is, splitting the cough sound in overlapping frames and computing features per frame). The frame length and the frame overlap were set to 50 ms and 50\%, respectively, and each frame was multiplied by a normalized Hamming window. Temporal features included zero-crossing rate, energy, energy entropy, and sound intensity, while spectral features comprised spectral centroid, spectral spread, spectral entropy, spectral flux, and 90\%-spectral roll-off. Additionally, Mel-frequency Cepstral Coefficients (MFCCs) and Filterbank Energies (Fbanks) were computed and appended to the previously mentioned feature set. For all spectral computations, a 1024-point Fast Fourier Transform (FFT) was used, while 40 filterbank coefficients and 13 MFFCs were selected per frame. Overall, 62 features per frame were computed. An example of the MFCCs computed is depicted in Fig.~\ref{fig:mfcc}.
Features were summarized using standard statistical approaches of mean, standard deviation, median, skewness, and kurtosis, along with the 1st, 99th, and the difference between the former two percentiles, resulting in 496 features per cough sound. A tSNE [15] plot of the resulting features in two dimensions in shown in Fig.~\ref{fig:tsne}.
Python and Scikit-Learn [12] were used for model implementation and training.
The given dataset is small and a simple train-test split might introduce bias while it leaves a significant amount of data out of training. We employed a stratified group 4-fold cross-validation (CV) scheme to verify model robustness and reduce performance bias (Figure 3). In general, K-fold cross-validation splits the entire dataset into k smaller sets. Each split is named a "fold". In each fold, a model is trained using K-1 of the folds as training data while the remaining part is used for testing. This strategy is repeated K times. When grouping is involved (group K-fold CV) it is ensured that the entire set of samples that belong to a specific group do not fall into both training and testing sets. Moreover, stratification (stratified K-fold CV) tries to preserve the distribution of classes in each split. In our study, a group is the set of cough sounds of a single participant and class is a binary variable (normal or abnormal X-ray result). Figure 3 illustrates this process.
Three machine learning models were developed, tested and compared including 1) logistic regression, 2) support vector machines and 3) multi-layer perceptron neural network [13]. Parameters of all models were tuned using five-fold stratified grouped cross-validation (CV). When necessary, features were standardized before being fed to a model.

\subsection{Machine Learning Models}
A brief description of the models follows next. Let us denote the acoustic features summarized in a P-dimensional vector x, where P=496 in our case.

\subsubsection{Logistic Regression}
Logistic regression (LR) [13] is a binary classifier that determines the probability of a patient having a normal or abnormal X-ray based on the acoustic features derived from his/her cough sound analysis. LR estimates the probability of abnormal X-ray, $P_{abn}(x)$, according to
\begin{equation}
    P_{abn}(x) = \frac{1}{1 + e^{-\theta^T x}}
\end{equation}
where $\theta$ is a $P-$dimensional parameter vector of the LR classifier. We also found that applying a regularization term based on the L2 norm is beneficial to the model. 

\subsubsection{Support Vector Machine}
Support Vector Machines (SVMs) [13] are classification models aiming to find the optimal hyperplane 
\begin{equation}
    w^T x + b = 0
\end{equation}
that separates classes by maximizing the margin between data points, known as support vectors. Here, w represents the perpendicular weights vector, x is a feature vector, and b is the bias term shifting the hyperplane. SVMs use a kernel function to map data into higher-dimensional spaces, enabling linear separation. 

\subsubsection{Multi-Layer Perceptron}

Multilayer Perceptrons (MLPs) [13] are a type of artificial neural network composed of multiple layers of interconnected nodes, each performing weighted summations followed by an activation function. The output of any layer in an MLP can be represented mathematically as 
\begin{equation}
A^{[l]}=\sigma\left(W^{[l]} A^{[l-1]} + b^{[l]} \right)
\end{equation}
where $A^{[l]}$ is the output vector of the $l^{th}$ layer, $\sigma(\cdot)$ is the activation function of the $l^{th}$ layer, $W^{[l]}$ denotes the weight matrix connecting layer $l-1$ to layer $l$, $A^{[l-1]}$ is the input vector from layer $l-1$ to layer $l$, and $b^{[l]}$ is the bias vector for the $l^{th}$ layer.

\section{Results}
\subsection{Enrollment}
This study was approved by the Royal Pune Independent Ethics Committee (RPIEC030723). We enrolled 137 patients from whom we recorded 967 coughs. The consent and cough collection process for each subject was brief and did not affect clinical operations. Of all radiographs, 36\% were interpreted as abnormal. The analysis of cough recordings showed that all three methods yielded models that could discriminate to some extent between normal and abnormal.

\subsection{Performance Metrics}
\begin{figure}
    \centering
    \includegraphics[scale=.6]{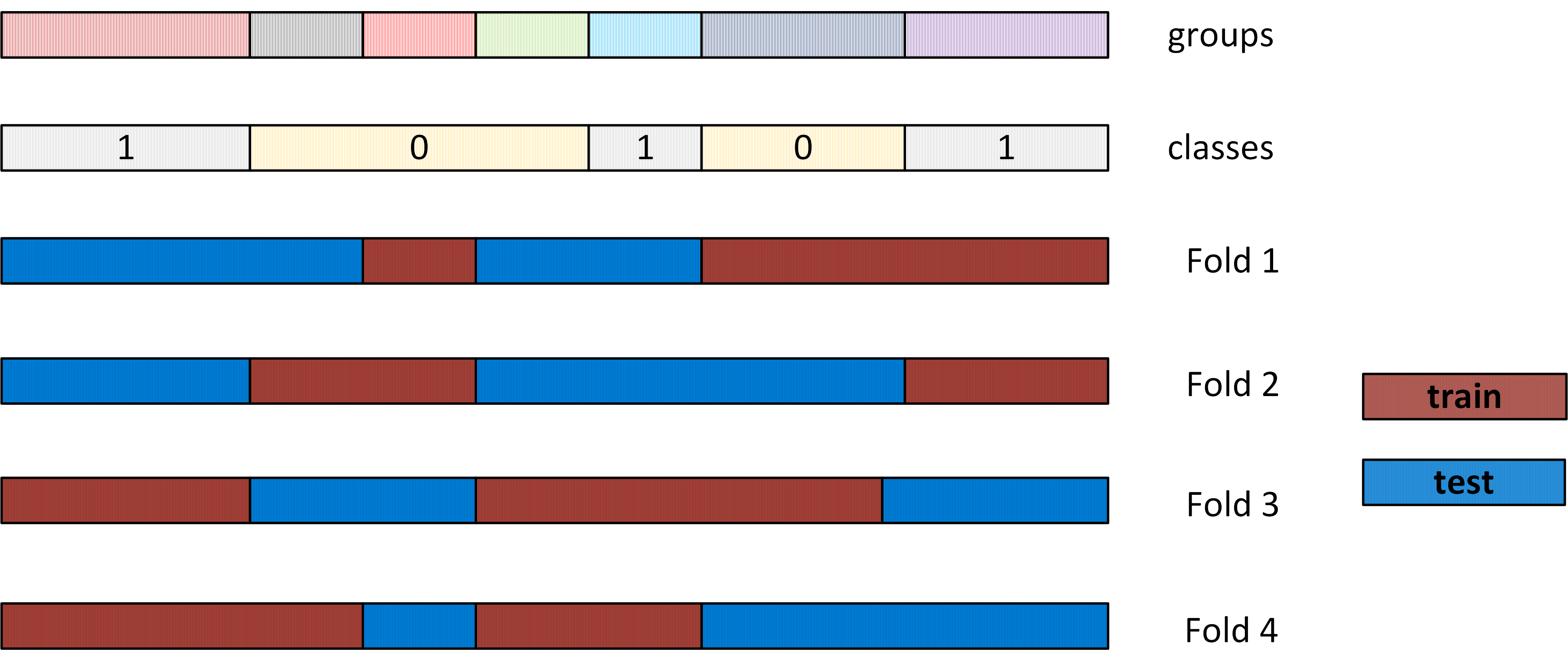}
    \caption{Graphical depiction of the stratified grouped four-fold cross validation. Groups are created as per the top most bar and AI models validated with each sub-sample (fold), avoiding any overlap between the train and test groups.}
    \label{fig:cv}
\end{figure}

We averaged the probabilities obtained from multiple cough sounds of the same patient in order to derive a per-patient probability. Given that each classifier outputs a probability, we denote TP, TN, FP, FN as the true positives, true negatives, false positives, and false negatives, respectively. We selected Sensitivity, defined as the true positive rate:
\begin{equation}
Sensitivity=\frac{TP}{TP+FN}
\end{equation}
Specificity, defined as the true negative rate:
\begin{equation}
Specificity=\frac{TN}{TN+FP}
\end{equation}
Precision, which is the positive predictive value:
\begin{equation}
Precision=\frac{TP}{TP+FP}
\end{equation}
F1-score, defined as the harmonic mean of the sensitivity and precision:
\begin{equation}
    F1 = \frac{2(Sensitivity \times Precision)}{Sensitivity + Precision}
\end{equation}
and the Area Under the Receiver Operating Characteristic curve (ROC-AUC) as the performance metrics, commonly used to show the diagnostic ability of the classifier and are generally acknowledged to be good measures of predictive accuracy. Specifically, ROCs for the four-fold stratified grouped CV methods are shown in Fig.~\ref{fig:cv}. 
Despite the small number of patients, the ROC-AUCs from the CV strategy range from 0.7 to 0.78, which is acceptable for triage and revealing that a simple LR model based on acoustic features can predict the X-ray status from cough sounds.

Table~\ref{tab:1} presents standard performance metrics for all classifiers as averages over all four folds.
\begin{table}[htb!]
\centering
\resizebox{0.55\columnwidth}{!}{%
\begin{tabular}{l|l|l|l|}
\cline{2-4}
 & LR & SVM & MLP-NN \\ \hline
\multicolumn{1}{|l|}{Sensitivity} & \bf{0.61 (0.07)} & 0.45 (0.10) & 0.43 (0.11) \\ \hline
\multicolumn{1}{|l|}{Specificity} & 0.77 (0.03) & \bf{0.84 (0.07)} & 0.79 (0.08) \\ \hline
\multicolumn{1}{|l|}{Precision} & 0.60 (0.05) & \bf{0.62 (0.18)} & 0.55 (0.14) \\ \hline
\multicolumn{1}{|l|}{ROC-AUC} & \bf{0.73 (0.02)} & 0.71 (0.04) & 0.69 (0.05) \\ \hline
\end{tabular}%
}
\caption{Performance metrics (mean, standard deviation) for all classifiers. LR, SVM, and MLP-NN stand for Logistic Regression, Support Vector Machines, and Multi-Layer Perceptron Neural Network, respectively. Best performance in bold.}\label{tab:1}
\end{table}

\section{Discussion}
We show that AI enabled algorithms can use cough sounds to predict which individuals presenting for chest radiographic examination will have a normal or abnormal results. We were able to collect coughs in the course of routine patient workflows, including consenting participants for this research study. Most importantly, the ROC-AUC was consistent with the performance characteristics of a screening or triage tool. 
These algorithms performed surprisingly well given the small size of the study but leave room for improvement. We anticipate such improvement will come from larger studies and the more sophisticated AI methods that larger data sets enable. Performance could be further augmented by incorporating additional readily available clinical information, such as fever, sputum on coughing, weight loss, or chest pain. Similar attempts to use cough classification for medical triage for tuberculosis improve significantly when they incorporate clinical signs such as the presence of productive cough, weight loss or hemoptysis [10, 14].
This pilot study suggests several areas for future research. From a technical perspective, other features can also be examined. Moreover, feature selection and dimensionality reduction can be applied to all models to reveal important features and speed up training. Also, more sophisticated models can be employed, such as ensemble methods [13] or deep learning models [16], when and if data is rich enough for such a task. From a clinical perspective, we only enrolled patients who clinicians had already decided to examine radiographically and thus likely had a higher pretest probability than would be found in patients presenting with respiratory complaints to primary care clinics. The months of January to March are cold in Northern India and pretext probability may also differ based on time of year.  Given the importance of testing triage technology in its intended use setting, additional studies will be needed to establish its performance over time in primary care clinics. The use of dedicated microphones in radiographic centers, while expeditious in this study, may limit the scalability and generalizability of cough screening. Adapting the technology for use on health care provider phones would increase the practicality of cough triage in primary or even community health care programs. Finally, this study was limited to adults, but given the challenges of managing pneumonia in children, expanding studies to the pediatric population has considerable potential public health impact. 

\section{Conclusion}
In summary, this pilot study suggests that the acoustic pattern of a solicited cough can help to identify which patients are most likely to have abnormal chest radiographs. If supported by larger appropriately designed studies, AI-enabled cough classification may become a routine tool used by front line healthcare personnel or even by the lay public. Such technology could improve the efficiency and efficacy of public health programs, particularly in low- and middle-income countries.

\section*{Acknowledgments}
This study was supported by funding and in-kind contributions from Hyfe and Qure.ai. The funders did not participate in the decision to submit these results for publication. ISGlobal acknowledges support from the Spanish Ministry of Science and Innovation through the “Centro de Excelencia Severo Ochoa 2019-2023” Program (CEX2018-000806-S), and support from the Generalitat de Catalunya through the CERCA program
Competing interests: SC, VS and SV are employees of Qure.ai., LJ, GPK, JB, PMS, JB, LJ and GK are employees of Hyfe, Inc., CCh has received consultancy fees from Hyfe, Inc., JB, PMS and CCh own equity from Hyfe, Inc. Data availability: The acoustic data and code used for this analysis are available at \url{https://github.com/hyfe-ai/QureAI-audio}.

\end{document}